\documentstyle[epsf,psfig]{mn}
\begin{document}
%\protect
%\onecolumn

\title[Galaxy-QSO correlation]
{Galaxy-QSO correlation induced by weak gravitational lensing arising from
  large-scale structure} 

\author[A. C. C. Guimar\~aes, C. van de Bruck and R. H. Brandenberger]
{Antonio C. C.\ Guimar\~aes$^1$, Carsten van de Bruck$^2$, and
  Robert H. Brandenberger$^1$ \\
$^1$Department of Physics, Brown University, Providence, RI 02912,
USA; guimar@het.brown.edu \\
$^2$DAMTP, Center for Mathematical Sciences, Wilberforce Road, CB3
0WA, Cambridge, U.K.
}

\date{4 October 2000}
\maketitle

%%%%%%%%%%%%%%%%%%%%%%%%%%%%%%%%%%%%%%%%%%%%%%%%%%%%%%%%%%%%%%%%%%%%%%%%%%%%%
%%%%%%%%%%%%%%%%%%%%%%%%%%%%    Abstract    %%%%%%%%%%%%%%%%%%%%%%%%%%%%%%%%%
%%%%%%%%%%%%%%%%%%%%%%%%%%%%%%%%%%%%%%%%%%%%%%%%%%%%%%%%%%%%%%%%%%%%%%%%%%%%%
\begin{abstract} 
Observational evidence shows that gravitational lensing induces an 
angular correlation between the distribution of galaxies and much more 
distant QSOs. 
We use weak gravitational lensing theory to calculate this angular
correlation, updating previous calculations and presenting new 
results exploring the dependence of the correlation on the large-scale
structure.  
We study the dependence of the predictions on a variety 
of cosmological models, such as cold dark matter models, mixed dark
matter models, and models based on quintessence. We also study
the dependence on the assumptions made about the nature of the
primordial fluctuation spectrum: adiabatic, isocurvature
and power spectra motivated by the cosmic string scenario are
investigated.  
Special attention is paid to the issue of galaxy biasing, which is
fully incorporated. 
We show that different mass power spectra imply distinct  
predictions for the angular correlation, and therefore the angular
correlation provides an extra
source of information about cosmological parameters and mechanisms of 
structure formation. 
We compare our results with observational data and discuss their
potential uses. 
In particular, it is suggested that the observational determination of the
galaxy-QSO correlation may be used to give an independent measurement
of the mass power spectrum.  

\end{abstract}

%\noindent BROWN-HET-1241, DAMTP-2000-77 

%%%%%%%%%%%%%%%%%%%%%%%%%%%%%%%%%%%%%%%%%%%%%%%%%%%%%%%%%%%%%%%%%%%%%%%%%%%%%
%%%%%%%%%%%%%%%%%%%%%%%%%%%%    Introduction    %%%%%%%%%%%%%%%%%%%%%%%%%%%%%
%%%%%%%%%%%%%%%%%%%%%%%%%%%%%%%%%%%%%%%%%%%%%%%%%%%%%%%%%%%%%%%%%%%%%%%%%%%%%

\section{INTRODUCTION}

According to general relativity, light from distant objects 
such as quasars is affected by the large-scale distribution 
of the intervening masses, and, consequently, these act as 
gravitational lenses. 
A gravitational lens enlarges the solid angle of the source and
conserves its surface brightness. So at the same time it brings to
view faint objects (magnification bias) and dilutes their population 
density. These two antagonistic effects can generate either correlation
or anti-correlation between the angular distribution of background and
foreground objects (see e.g. Schneider, Ehlers \& Falco 1992).

Several studies have explored the correlation between 
QSO's and galaxies at various wavelengths. For example, 
Bartelmann \& Schneider (1994) found a statistically significant 
correlation between 1Jy sources and IRAS galaxies for quasars 
at high redshifts. 
Also, Ben\'{\i}tez \& Mart\'{\i}nez-Gonz\'alez (1997) 
reported statistically significant correlations between COSMOS/UKST
galaxies and PKS radio quasars, 
and Ben\'{\i}tez, Sanz \& Mart\'{\i}nez-Gonz\'alez (2001) found
positive correlations using radio-loud quasars from the 1Jy and
Half-Jansky samples.   
These results were obtained for small 
angles on the sky (of the order of 10 arcmin). Recently, Williams 
\& Irwin (1998) investigated the correlation between QSO's from the 
LBQS catalog and galaxies from the APM survey and detected a 
significant positive angular correlation, ranging from scales of 10 arcmin 
to over a degree. For a more extended review of observational
data see Ben\'{\i}tez et al. (2001), 
Bartelmann \& Schneider (2001), and Norman \& Willians (2000).

An analytical theory of the galaxy-QSO correlation due to weak 
gravitational lensing was introduced by Bartelmann (1995), 
Dolag \& Bartelmann (1997), hereafter B95 and DB97, and Sanz,
Mart\'{\i}nez-Gonz\'alez \& Ben\'{\i}tez (1997), having as base the work by
Kaiser (1992).  
We will review the theory briefly in Section 2. 
The predicted correlations between galaxies and QSO's agree reasonably 
well with some observations for angles below 10 arcmin. 
However, Williams \& Irwin (1998) claim that for larger angular
scales theory and observations differ by one order of magnitude for standard 
cosmological models such as cold dark matter ($SCDM$) and cold dark
matter with a cosmological constant ($\Lambda CDM$). 

Although this conclusion has to be tested with future observations, 
it is of importance to investigate the predictions of the 
galaxy-QSO correlation for a large class of cosmologies. 
The effect of bias, matter content and current expansion rate has
to be investigated in detail. This will be the aim of this paper, 
where we keep the assumption that the weak lensing approximation 
is valid. Effects of higher order terms are not investigate here.
These contributions where estimated recently by Williams (2000), 
and were shown to increase the amplitude by less than 10 per cent 
for nearby large-scale coherent structures. 
Our investigations show that the observations of correlations between
QSOs and galaxies on large angular scales can provide independent
information relevant to structure formation understanding, and are somewhat
sensitive to the values of a set of cosmological parameters.

The paper is organized as follows. 
In Section 2 we review the theoretical formalism used.  
In Section 3 we present the cosmological models we consider in our 
investigation and study their predictions. The models 
include mixed dark matter and quintessence models, as well as isocurvature  
and cosmic string based models of structure formation. 
We analyze in detail the effect of several cosmological
parameters on the galaxy-QSO correlation in a flat universe with
cosmological constant and initial adiabatic fluctuations.
We discuss our findings and some observational data in Section 4. 

% \newpage

%%%%%%%%%%%%%%%%%%%%%%%%%%%%%%%%%%%%%%%%%%%%%%%%%%%%%%%%%%%%%%%%%%%%%%%%%%%%%
%%%%%%%%%%%%%%%%%%%   Theoretical Framework    %%%%%%%%%%%%%%%%%%%%%%%%%%%%%%
%%%%%%%%%%%%%%%%%%%%%%%%%%%%%%%%%%%%%%%%%%%%%%%%%%%%%%%%%%%%%%%%%%%%%%%%%%%%%

\section{THEORETICAL FRAMEWORK}
\label{Framework}
In this section we review the theory of the correlation between 
galaxies and QSOs based on weak gravitational lensing. 
We follow closely Bartelmann (1995) and Dolag \& 
Bartelmann (1997). However, we extend the theory by 
allowing for a scale- and time-dependent bias, and by using an improved mass
power spectrum (e.g. by taking into account the baryon density and
properly evolving non-linear density fields). 

\subsection{Angular correlation definition and calculation}
We define the angular correlation function $\xi _{GQ}(\phi)$ 
between galaxies and high-redshift QSO's as 
\begin{equation}
\xi_{GQ}(\phi) \equiv 
\left< \left[ \frac {n_Q(\vec{\theta})}{\bar{n}_Q} -1 \right]
\left[ \frac {n_G( \vec{\theta}+ \vec{\phi})}{\bar{n}_G} -1 \right]  
\right>  \, ,
\end{equation}
where $n_Q$ and $n_G$ are the QSO and galaxy densities (a bar over a quantity 
indicates its mean value), and $\left< \; \right>$ represents the average
over $\vec{\theta}$ and the direction of $\vec{\phi}$ (but not its modulus).
Assuming weak gravitational lensing (i.e. that the magnification 
field contrast $\delta\mu$ is small, $|\delta\mu| \ll 1$), and 
a brightness distribution of the cumulative quasar number density as a
function of flux $S$ of the form $n_Q(>S)\propto S^{-s}$, one can show
(B95) that 
\begin{equation}
\xi_{GQ}(\phi) = (s-1) \, \langle \delta \mu (\vec{\theta})
\delta_{gal}(\vec{\theta}+ \vec{\phi}) \rangle  \, ,
\label{xiQGdef}
\end{equation}
where $\delta_{gal}$ is the galaxy density contrast.

The magnification fluctuations $\delta\mu$ derive from the 
inhomogeneities in the mass distribution, since these create a
gravitational potential topography which deflects light, 
and therefore generates a shear field. The calculation of this shear
field (which assumes a statistically homogeneous and isotropic density
field and a Newtonian weak field approximation for gravity), 
and the results involving the power spectra of projections of
three-dimensional fields, obtained by Kaiser (1992), allows one
to derive the expression 
[equation (12) of DB97 generalized for a scale and time dependent bias 
factor $b(k,w)$ (Ben\'{\i}tez \& Sanz 1999)] 
\begin{eqnarray}
\frac{\xi _{GQ}(\phi)}{s-1} & = &
{\displaystyle \frac {3 \Omega_{m}}{2 \pi (c/H_0)^2 }} 
{\displaystyle \int _{0}^{w _{\infty} }} dw 
{\displaystyle \frac { W_G(w) \, G_Q(w) } {a(w)}}  \nonumber \\
& &  \times
{\displaystyle \int _{0}^{\infty }} dk \, k \, b(k,w) 
\, P_{\delta}(k,w )\,\mathrm{J_0} 
[f_K(w )k\phi ]  \, , 
\label{main}
\end{eqnarray}
where 
$w$ is the comoving distance which here parameterizes time
($w_{\infty}$ represents a redshift $z=\infty$), 
and $k$ is the wavenumber of the density contrast in a plane wave
expansion;
$\Omega_{m}$ is the matter density parameter;
$P_{\delta}(k,w)$ is the time evolved mass power spectrum;
$J_0$ is the zeroth-order Bessel function of first kind;
and $f_K(w)$ is the curvature-dependent radial distance ($=w$ for a flat 
universe).  

The scale factor, $a=1/(1+z)$, is determined by solving the following
integral equation for $a$
\begin{equation}
w(a) =  
\int _{a}^1 {\displaystyle \frac {da^{\prime}}
{a^{\prime 2} H(a^{\prime})/H_0 }} 
\, 
\label{w(a)}
\end{equation}
where 
$H(a)= H_0 \, [ \Omega_{\Lambda} (1-a^{-2}) + 
\Omega_m(a^{-3}-a^{-2})+ a^{-2} ]^{1/2}$.
For an Einstein-de Sitter Universe ($\Omega_m =1$ and $\Omega_{\Lambda}=0$)
the last equation can be easily solved, implying
$ a(w) = \left( 1 - w /2 \right)^2 $,
$w$ being measured in units of $c/H_0$. For a general cosmology 
equation (\ref{w(a)}) has to be solved numerically.

The information about galaxy and QSO redshift distribution enters through 
the weight functions $W_G(w)$ and $G_Q(w)$, respectively.
We take $W_G(w)$ from Kaiser (1992) with parameters $\alpha=1$, 
$\beta=4$ and a mean galaxy redshift $z_G=0.2$, that is just a
distribution sharply peaked at $z_G$, which mimics the
APM survey redshift distribution well. For the QSO redshift distribution we
adopt a simple linear ramp function which starts at a high value at a
minimum redshift $z_0=0.3$, and falls to zero at $z=3.5$. This also
well mimics QSO surveys such as LBQS and PKS. From this distribution the
QSO weight function $G_Q(w)$ is calculated following DB97. Note
however that the exact shape of these distributions is not crucial
for our results, and its role is already discussed in B95.

%%%%%%%%%%%%%%%%%%%%%%%%%%%%%%%%%%%%%%%%%%%%%%%%%%%%%%%%%%%%%%%%%%%%%%%%
\subsection{Mass power spectrum}

We define the linear mass power spectrum as usual
\begin{equation}
\Delta^2(k,w) \equiv 
\frac{k^3}{2 \pi^2} P_{\delta}(k,w) = A 
\left(\frac{c k}{H_0} \right)^{3+n} T^2(k,w) D^2(w) \, ,
\end{equation} 
where $A$ is the normalization factor\footnotemark
\footnotetext{$A= \delta_H$ if we normalize
to COBE (Bunn \& White 1997). Another possibility is to 
normalize to the cluster abundance (Viana \& Liddle 1999), 
which would be in principle the most
appropriate for our calculations, because it normalizes the power
spectrum on scales that are relevant to weak lensing, while a
normalization to the cosmic microwave background anisotropies does
the normalization on much larger scales.}
; $n$ is the initial power spectrum index (for a Harrison-Zel'dovich 
spectrum $n=1$); $T(k,w)$ is the transfer function; and $D(w)$ 
is the growth function which we define as 
\begin{equation}
D(w) = a \frac{g(a)}{g(1)} \, ,
\label{growth}
\end{equation}
where $a=a(w)$, and $g(a)$ is the linear growth suppression 
factor which is well approximated by (Carroll, Press \& Turner 1992; 
Lahav et al. 1991) 
\begin{eqnarray}
g(a) & = & \frac 5{2}  \Omega_m(a) \left[ \Omega_m^{4/7}(a)
- \Omega_{\Lambda}(a) \vphantom{\frac 1{2}} \right.  \nonumber \\
& & + \left. \left( 1+ \frac 1{2} \Omega_m(a) \right)
\left( 1+ \frac 1{70} \Omega_{\Lambda}(a) \right) \right] ^{-1} \, ,
\end{eqnarray}
\begin{equation}
\Omega_m(a) = \frac {\Omega_m}{a^3} \left( \frac{H_0}{H} \right)^2 \, ,
\end{equation}
\begin{equation}
\Omega_{\Lambda}(a) = {\Omega_{\Lambda}} \left( \frac{H_0}{H} \right)^2 \, .
\end{equation}
For an Einstein-de Sitter Universe $g(a)=1$ (no suppression of
gravitational clustering), and the growth function 
reduces to the scale factor, $D(w) = a(w)$.

For CDM we use the transfer functions given by Bardeen et al. (1986),
calculated with the shape parameter by Sugiyama (1995),
$\Gamma =  \Omega_m h \exp [-\Omega_b (1+\sqrt{2 h}/\Omega_m)] $,
where $\Omega_b$ is the baryon energy density.

A realistic model for the mass power spectrum has to consider non-linear 
effects in the evolution of the density fields, important 
at small scales. One way to do this is by computer-intensive simulations, 
another is to use an approximate analytical form such as a scaling equation 
mapping the linear spectra to the non-linear one (Hamilton et
al. 1991, Scranton \& Dodelson 2000)
\begin{equation}
\Delta^2_{NL}(k_{NL},w) = f_{NL} \left(  \Delta^2_{L}(k_L,w) 
\right) 
\end{equation}
\begin{equation}
k_{NL} = k_L \left[1+\Delta^2_{NL}(k_{NL},w) \right]^{1/3} \, .
\end{equation}
Expressions for the function $f_{NL}(x)$ are given by 
Peacock \& Dodds (1996) and Ma (1998), hereafter PD96 and Ma98. 
In the linear regime ($x\ll1$) $f_{NL}(x)\approx x$, during the stable
clustering regime ($x\gg1$) $f_{NL}(x)\propto x^{3/2}$, and in the 
intermediate region ($x\approx1$) $f_{NL}(x)$ is extracted from N-body
simulations.

%%%%%%%%%%%%%%%%%%%%%%%%%%%%%%%%%%%%%%%%%%%%%%%%%%%%%%%%%%%%%%%%%%%%%%%
\subsection{Galaxy biasing}

The simplest bias relation one can assume is linear and deterministic, 
so that one could define the bias as
\begin{equation}
b = \frac{(\sigma_8)_{gal}}{(\sigma_8)_{mass}} \, ,
\label{lbias}
\end{equation}
where 
\begin{equation}
\sigma ^2(R) = \int^{\infty}_0 W^2(kR) \Delta^2(k) \frac {dk}{k} \, ,
\end{equation}
and
$ W(x)=3 \left[ \sin(x) - x \cos(x) \right] / x^3 $. 
Evaluated at $R=8h^{-1}Mpc$, this gives $\sigma_8$. The subscripts 
`gal' and `mass' refer to the galaxy and mass power spectra. 

However, the linear deterministic bias is a too simplistic model 
(Dekel \& Lahav 1999).
The most general biasing between galaxy and mass density contrasts that 
one can imagine is a non-linear stochastic bias which can be time- and 
scale-dependent:
\begin{equation}
\delta_{gal} = b(\delta) \delta + \epsilon \, ,
\label{bias}
\end{equation}
where $\epsilon$ is a random field.

From equation (\ref{xiQGdef}) we get that
\begin{equation}
\frac{\xi_{QG}(\phi)}{(s-1)} = \left< \delta \mu ({\vec{\theta}})
 b[\delta({ \vec{\theta}}+{ \vec{\phi}})] \delta({\vec{\theta}}+
{\vec{\phi}}) \right> 
+  \left< \delta \mu ({\vec{\theta}})  \epsilon({\vec{\theta}}+
{\vec{\phi}}) \right> \, .
\label{bias-xi}
\end{equation}
If we make the reasonable assumptions that $\epsilon$ is independent
of $\delta \mu $ and has zero mean, then the last term on the right is
clearly zero\footnotemark
\footnotetext{While this is certainly true for averages over large
  angular regions, for patches smaller than the typical
  scales of variation of the fields involved this term could
  contribute because of statistical variance.}
, so a stochastic component of the biasing does not alter
the expected value for the galaxy-QSO correlation. 

To be most general, we further consider a scale- and time-dependent
bias factor, which we assume can be written as $b(k,w) = b_s(k) b_t(w)$.
The scale-dependent part of the bias is defined as 
\begin{equation}
b_s(k) = \sqrt{\frac{P_{gal}(k)}{P_{\delta}(k)}} \, ,
\label{b_k}
\end{equation}
where $P_{gal}(k)$ is an empirical galaxy power spectrum. 
We use (unless indicated) the spectrum obtained from the APM survey 
(Gazta\~naga \& Baugh 1998).
The time-dependent part of the bias can be modeled as 
(Tegmark \& Peebles 1998)
\begin{equation}
b_t(w) = \frac{\sqrt{(1-D)^2 - 2(1-D)r_0 b_0 + b_0^2}}{b_0 D} \, ,
\end{equation}
where $b_0$ is the linear bias factor today (equation [\ref {lbias}]), 
$r_0$ is the dimensionless correlation coefficient between the 
distributions of mass and galaxies, and $D=D(w)$ (equation [\ref{growth}]). 
However, we found that the time dependence is negligible (even
considering a galaxy population at a higher redshift and a strong 
time dependence of the bias the effect is very small), so for simplicity 
one can neglect it.

%%%%%%%%%%%%%%%%%%%%%%%%%%%%%%%%%%%%%%%%%%%%%%%%%%%%%%%%%%%%%%%%%%%%%%%%%%%%%
%%%%%%%%%%%%%%%%%%%%%%%%%%%%%%     RESULTS   %%%%%%%%%%%%%%%%%%%%%%%%%%%%%%%%
%%%%%%%%%%%%%%%%%%%%%%%%%%%%%%%%%%%%%%%%%%%%%%%%%%%%%%%%%%%%%%%%%%%%%%%%%%%%%

\section{RESULTS}
\label{RESULTS}
In this section (except for Section 3.4) we explore the dependence of the
galaxy-QSO angular correlation on several cosmological parameters, assuming 
structure formation from initial adiabatic fluctuations. 

\subsection{Three historical models}

\begin{figure}
%\centering 
%\leavevmode 
%\gdef\eps@scaling{0.7}
%\includegraphics[width={\eps@scaling\columnwidth}]{3cosmos-PS-bias.eps}
%\includegraphics[width={\eps@scaling\columnwidth}]{3cosmos-GQC.eps}
\epsfxsize=3.6in
\epsfbox{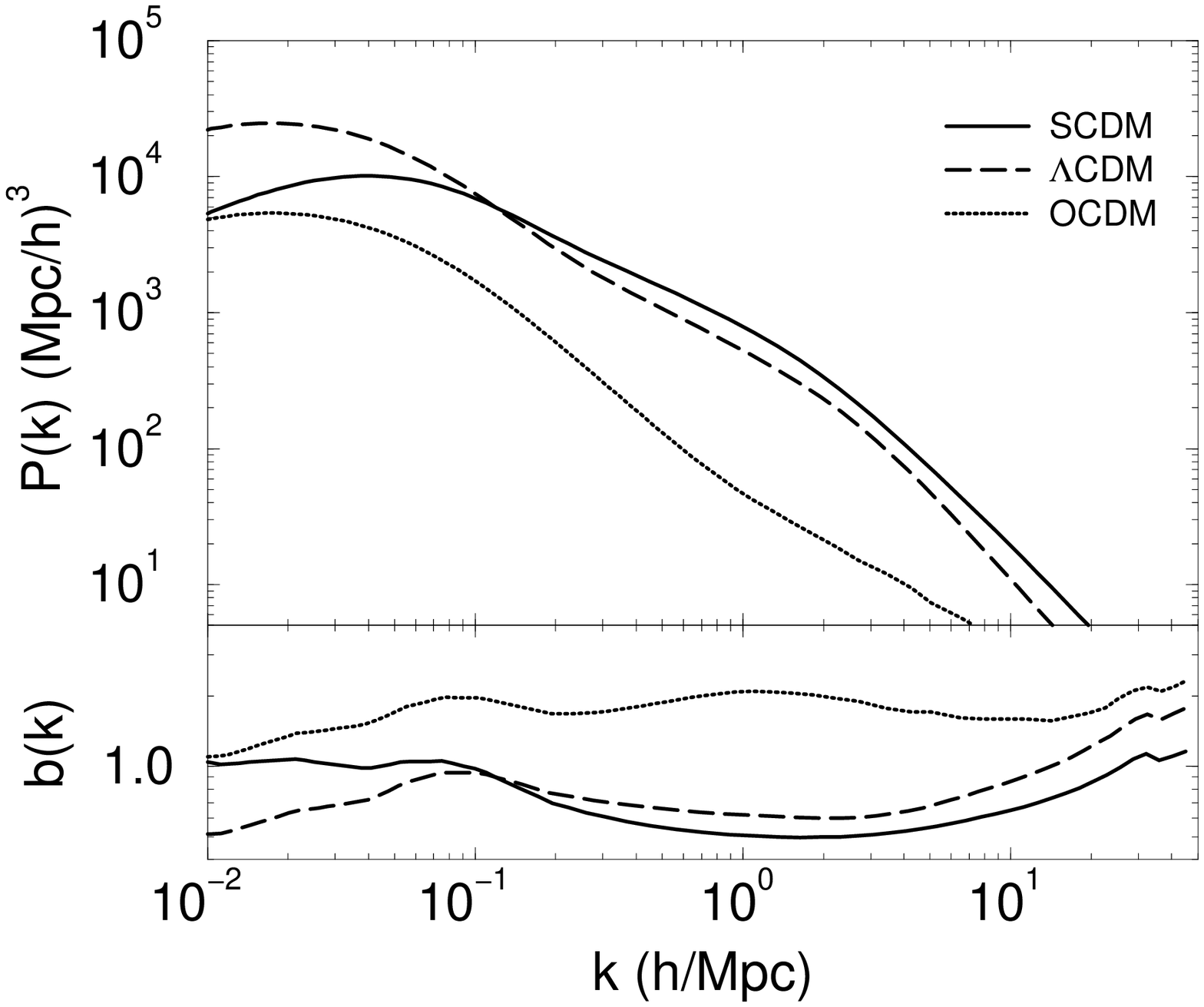}
\epsfxsize=3.6in
\epsfbox{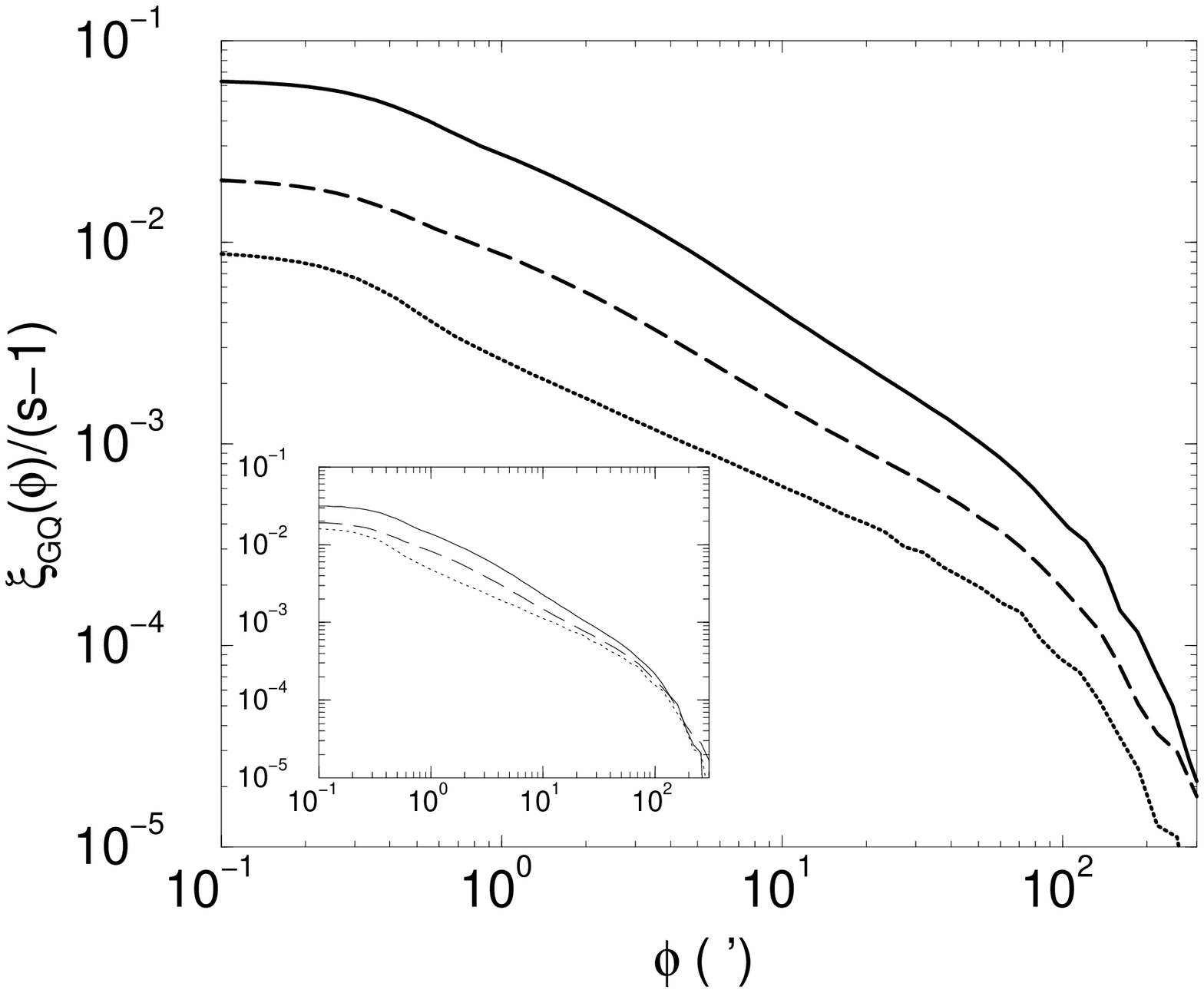}
\caption{Three cosmological models.
Non-Linear mass power spectrum (top graph), bias (central graph), and
angular correlation $\xi_{GQ}(\phi)/(s-1)$ (bottom graph). 
The internal plot is the result for the mass power spectrum normalized
to the cluster abundance. 
Solid lines are for $SCDM$, $\Omega_m=1$, $h=0.5$ ($\sigma_8=1.1$); 
dashed ones for $\Lambda CDM$, $\Omega_m=0.3$, $\Omega_{\Lambda}=0.7$,
$h=0.7$ ($\sigma_8=1.0$); and
dotted ones for $OCDM$, $\Omega_m=0.3$, $\Omega_{\Lambda}=0$, $h=0.7$ 
($\sigma_8=0.46$).
Other parameters are 
$\Omega_b=0.019/h^2$, $n=1$, non-linear approximation by PD96.}
\label{3cosmologies}
\end{figure}

Fig. \ref{3cosmologies} shows our results for three cosmological models that 
were in fashion at some point in history, the `standard' cold dark matter
($SCDM$: $\Omega_m=1$, $h=0.5$), a flat universe with a cosmological
constant ($\Lambda CDM$: $\Omega_m=0.3$, $\Omega_{\Lambda}=0.7$, $h=0.7$), 
and an open universe ($OCDM$: $\Omega_m=0.3$, $\Omega_{\Lambda}=0$, $h=0.7$).
It uses the COBE normalized mass power spectrum for these models, 
the necessary bias to match the APM galaxy power spectrum (equation [\ref{b_k}]),
and the calculated galaxy-QSO angular correlation (equation [\ref{main}]).
The three cosmological models are clearly differentiated by the 
predicted $\xi_{GQ}(\phi)$. 

A normalization of the power spectrum to the cluster abundance simply
rescales our results for $\xi_{GQ}(\phi)$ by 
$(\sigma_8)_{cluster}/(\sigma_8)_{COBE}$. The resulting correlations
decrease for $SCDM$, increase for $OCDM$, and remain the same for 
$\Lambda CDM$, making the differentiation among the models less evident.

The latest CMB observation (Bernardis et al. 2000, and Hanany et
al. 2000) point to a flat universe, as favored by most
inflationary models. 
The result of these observations, and the suggestion of the
existence of a dark energy component by supernova observations,
strongly induce one to take 
$\Omega \equiv \Omega_m + \Omega_{\Lambda} = 1$.  
This flat universe with low matter density (most of it coming from
some sort of cold dark matter), dark energy (cosmological constant
or quintessence), and adiabatic fluctuations with an initial spectrum 
nearly scale-free, constitutes the most popular cosmological model today. 
We will concentrate on this model, searching for the effects of 
variations of the cosmological parameters around the most likely values.

\subsection{Adiabatic flat universe with $\Lambda$}

The matter density parameter $\Omega_m$ enters the calculation of 
$\xi_{GQ}(\phi)$ at several levels. It is a multiplicative factor of the 
whole expression in equation (\ref{main}), it determines the general geometric 
scaling of the problem through the scale factor, and it also enters in the  
mass power spectrum (and consequently in the bias, too) through the shape 
parameter, the growth function, and normalization.

\begin{figure}
\centering 
\epsfxsize=3.6in
\epsfbox{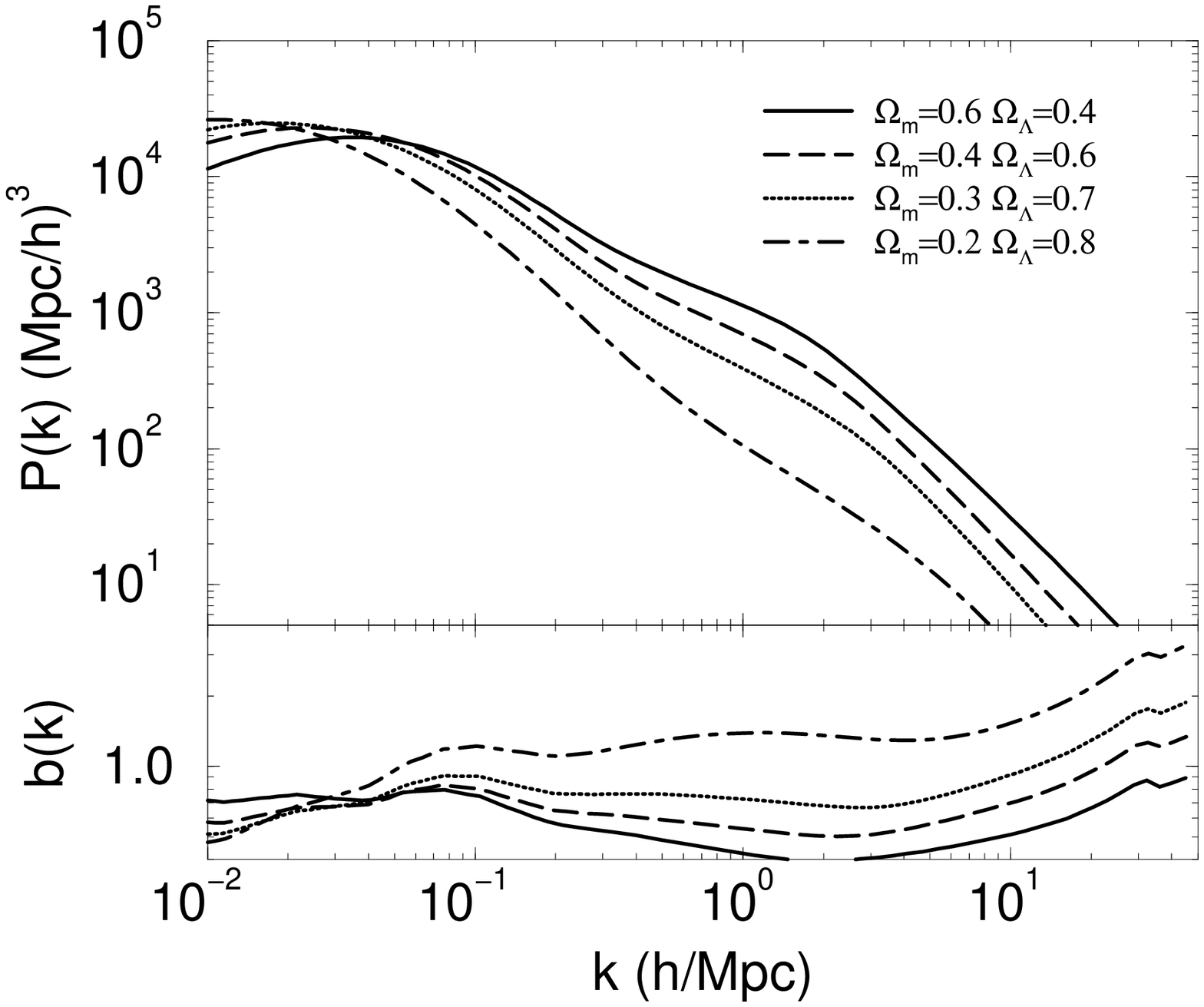}
\epsfxsize=3.6in
\epsfbox{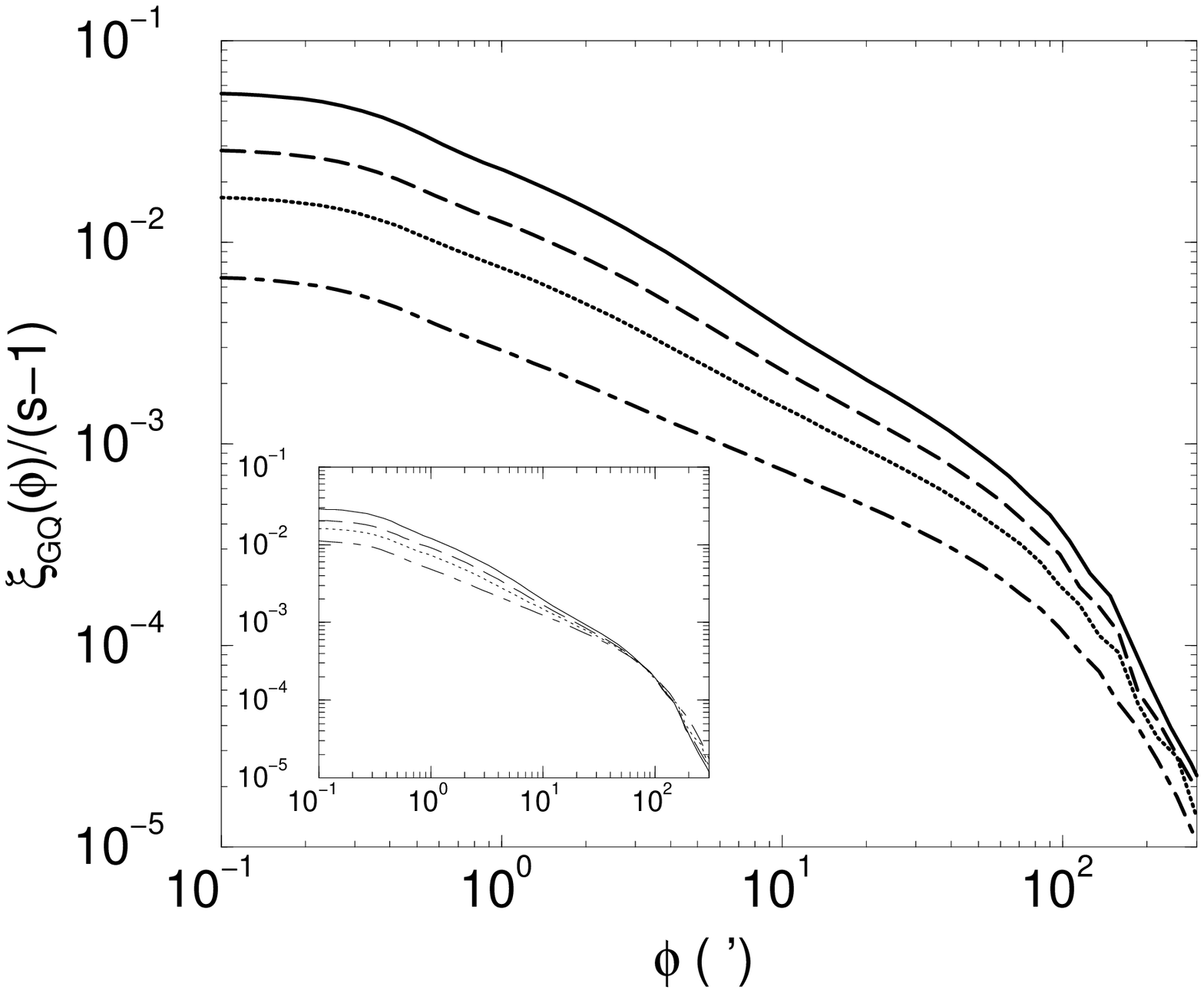}
\caption{Dependence on matter density in a flat universe with
  cosmological constant ($\Omega_m + \Omega_{\Lambda}=1$). 
Solid lines are for $\Omega_m=0.6$ ($\sigma_8=1.4$); 
dashed ones for $\Omega_m=0.4$ ($\sigma_8=1.2$);
dotted ones for $\Omega_m=0.3$ ($\sigma_8=1.0$); and
dot-dashed ones for $\Omega_m=0.2$ ($\sigma_8=0.72$).
Other parameters are 
$\Omega_b=0.019/h^2$, $h=0.7$, $n=1$, non-linear approximation by Ma98.}
\label{flatV-m}
\end{figure}

We investigate the dependence of the galaxy-QSO correlation on the matter 
content of a flat universe in Fig. \ref{flatV-m}. As would be expected, 
a higher matter content generates more power of the mass 
spectrum on small scales, which results in a higher correlation between galaxies and QSO's (mainly at small angles).

Recent experiments give extra support to the possibility
that neutrinos are massive. This necessarily implies a hot dark matter (HDM)
component to the matter content of the universe, because theory predicts a
cosmic neutrino background of temperature 1.95 K. 
A pure HDM universe is ruled out, because it would not have the
observed amount of structure on small scales, as they are suppressed by 
free streaming. However, a mixed dark matter scenario (MDM) with a
small fraction of HDM is possible.

\begin{figure}
\centering 
\epsfxsize=3.6in
\epsfbox{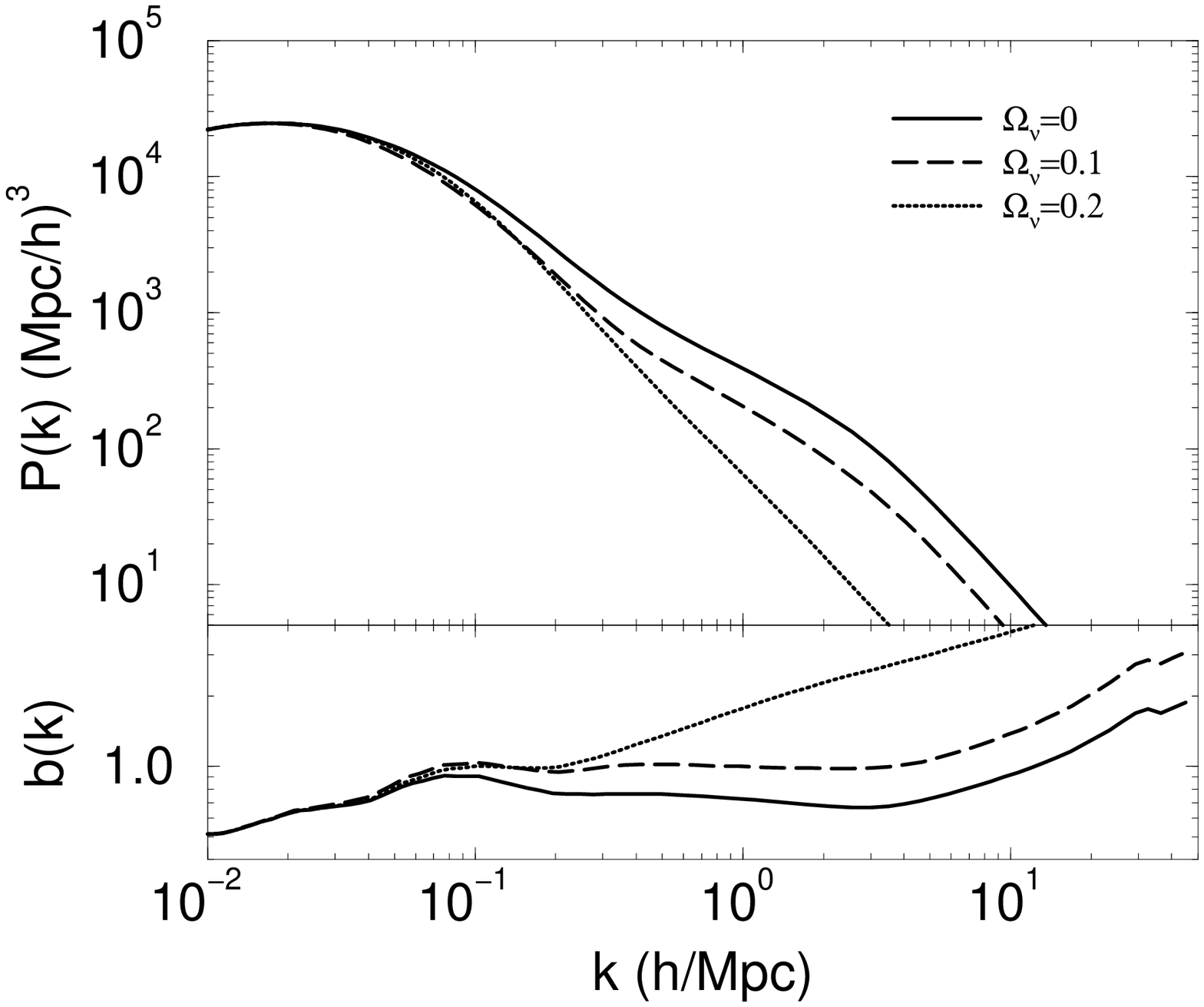}
\epsfxsize=3.6in
\epsfbox{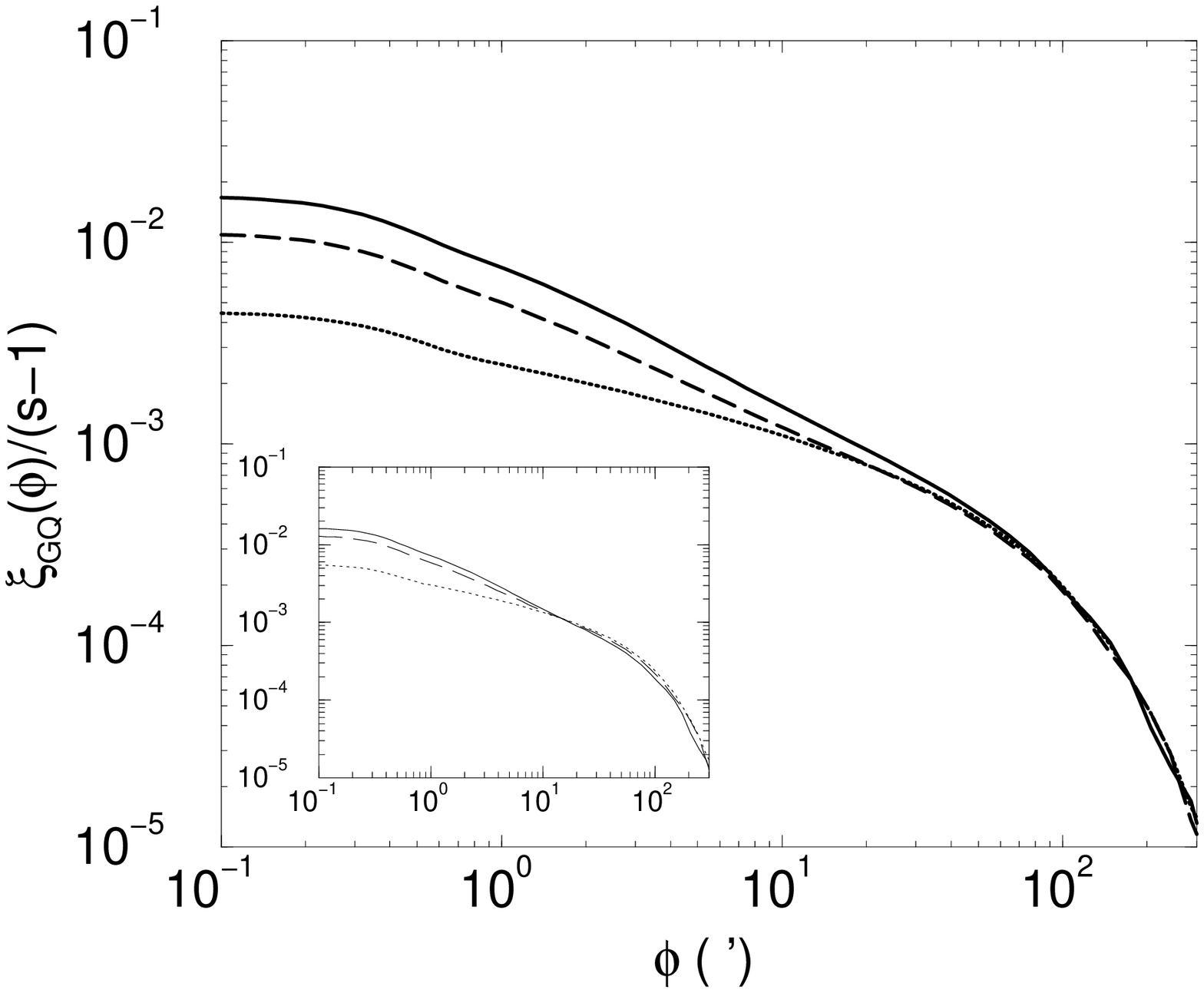}
\caption{Dependence on hot dark matter density.
Solid lines are for $\Omega_{\nu}=0$ ($\sigma_8=1.0$); 
dashed ones for $\Omega_{\nu}=0.1$ ($\sigma_8=0.84$); and
dotted ones for $\Omega_{\nu}=0.2$ ($\sigma_8=0.81$).
Other parameters are  
$ (\Omega_m, \Omega_{\Lambda},  \Omega_b) 
= (     0.3,              0.7, 0.019/h^2)$, 
$h=0.7$, $n=1$, non-linear approximation by Ma98.}
\label{MDM}
\end{figure}

We calculate $\xi_{GQ}(\phi)$ in a MDM universe using the linear mass power 
spectrum derived by Ma (1996) and obtain the full non-linear time evolved 
spectrum using the prescription by Ma (1998), which properly takes into 
account the time evolution of the hot and cold dark matter
components (the PD96 prescription does not give the correct result in this case). 
Fig. \ref{MDM} shows our results. The larger the fraction of 
HDM over CDM, the larger the suppression of structures on small scales. 
A large neutrino energy density $\Omega_{\nu}$ results in less power
at large $k$. The consequence on $\xi_{GQ}(\phi)$
is a lower correlation at small angles (and a flatter general slope). 

\begin{figure}
\centering 
\epsfxsize=3.6in
\epsfbox{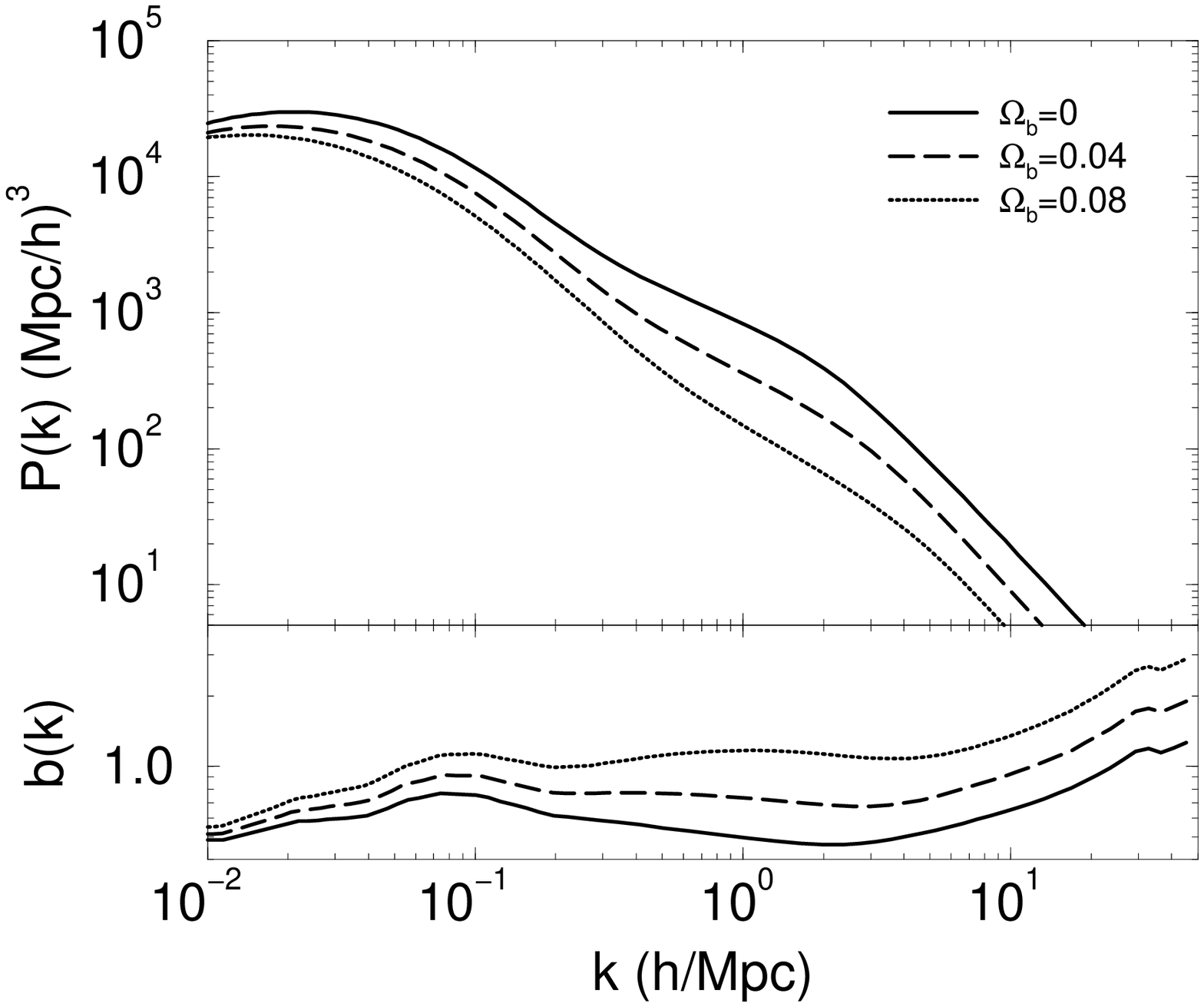}
\epsfxsize=3.6in
\epsfbox{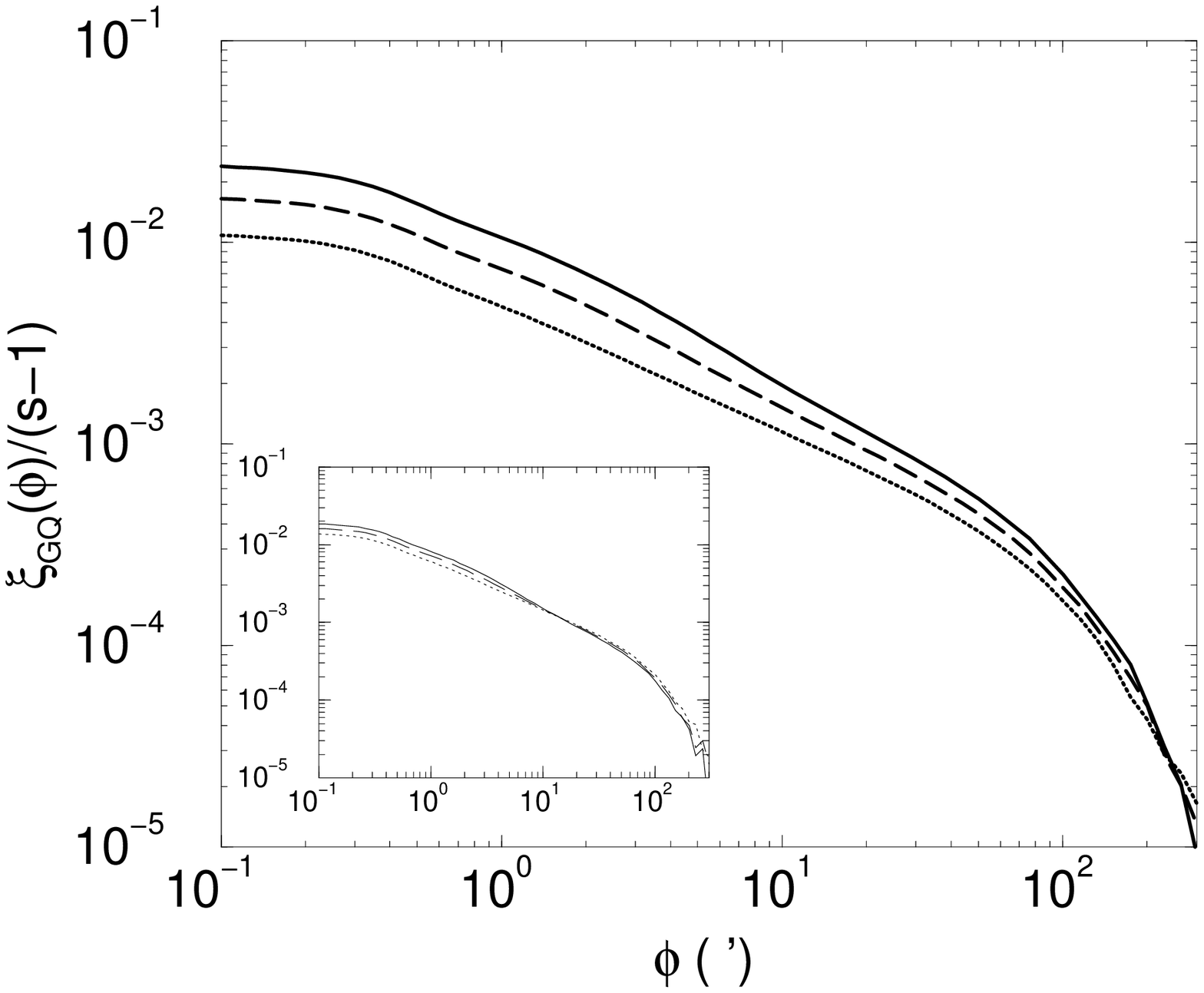}
\caption{Dependence on baryon density.
Solid lines are for $\Omega_{b}=0$ ($\sigma_8=1.3$); 
dashed ones for $\Omega_{b}=0.04$ ($\sigma_8=1.0$); and
dotted ones for $\Omega_{\nu}=0.08$ ($\sigma_8=0.79$).
Other parameters are 
$ \Omega_m=0.3$, $\Omega_{\Lambda}=0.7$,  
$h=0.7$, $n=1$, non-linear approximation by Ma98.}
\label{baryon}
\end{figure}

The baryon density of the universe is fairly well determined 
(Burles et al. 1999) and represents a small fraction of the total matter. 
In our calculations $\Omega_b$ enters through the shape parameter 
$\Gamma$ of the transfer function. This is just an approximation, because it 
does not account for baryonic oscillations (Eisenstein \& Hu 1998),
but it is good enough for low values of $\Omega_b$. 
We studied the influence of a baryonic content varying from 0 to 8 per centof the
critical density on the galaxy-QSO correlation (see
Fig. \ref{baryon}). A higher baryon density implies less power on
small scales for the mass spectrum and a lower value of
$\xi_{GQ}(\phi)$ on small angles. 

\begin{figure}
\centering 
\epsfxsize=3.6in
\epsfbox{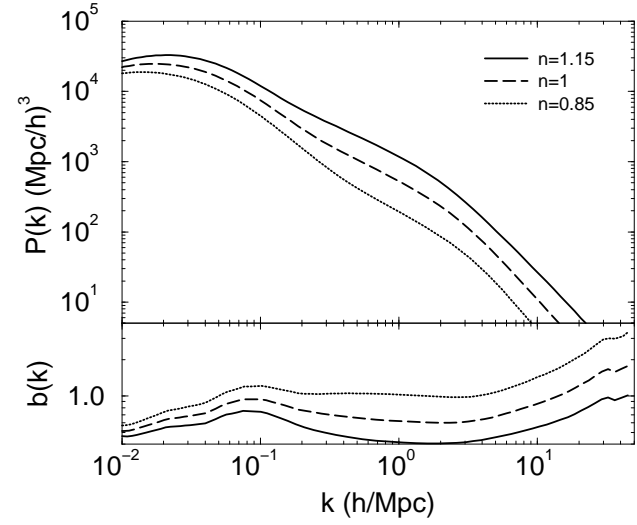}
\epsfxsize=3.6in
\epsfbox{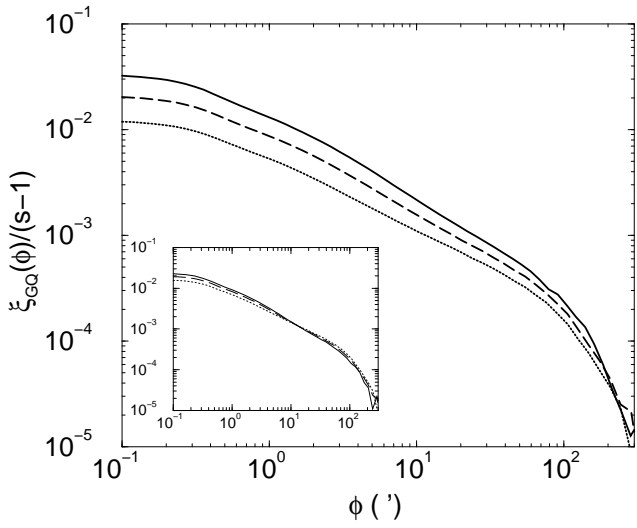}
\caption{Dependence on the primordial spectral index.
Solid lines are for $n=1.15$ ($\sigma_8=1.4$); 
dashed ones for $n=1$ ($\sigma_8=1.0$); and
dotted ones for $n=0.85$ ($\sigma_8=0.75$).
Other parameters are 
$ (\Omega_m, \Omega_{\Lambda},  \Omega_b) 
= (     0.3,              0.7, 0.019/h^2)$, 
$h=0.7$, non-linear approximation by PD96.}
\label{n}
\end{figure}

Inflation favors an initial power index $n \approx 1$ for the spectrum 
of density fluctuations, but some degree of `tilting' is possible
depending of the inflationary theory.
We look at the effect of modifying the initial power spectrum 
index by simple tilting at Fig. \ref{n}. A higher $n$ implies a
steeper $\xi_{GQ}(\phi)$, since it induces more power on small scales.

The effect of a Hubble parameter in the range $0.5<h<0.8$ on the 
predicted $\xi_{GQ}(\phi)$ was also considered. A higher value for $h$
implies a higher correlation, as also previously shown in DB97. 

\subsection{Quintessence}

One alternative for the cosmological constant as dark energy is a 
dynamical and spatially homogeneous (on cluster scales) field, or, as
it has been called, quintessence (see e.g. Wetterich 1988, Ratra and
Peebles 1988). 
Such a field has negative pressure 
and an equation of state $p=\omega_Q\rho$ with coefficient in the range 
$-1\leq\omega_Q<0$. 
We implement a quintessence model ($QCDM$) with constant $\omega_Q$ by 
generalizing the time dependent Hubble expansion rate to 
\begin{equation}
H(a)= H_0 \, [ \Omega_Q (a^{-3(\omega_Q+1)}-a^{-2}) + 
\Omega_m(a^{-3}-a^{-2}) + a^{-2} ]^{1/2} \, ,
\end{equation} 
and by using the results of Ma et al. (1999) for the construction of
the fully evolved non-linear mass power spectrum.
Note that $\omega_Q=-1$ reduces everything to the case of a cosmological
constant.

\begin{figure}
\centering 
\epsfxsize=3.6in
\epsfbox{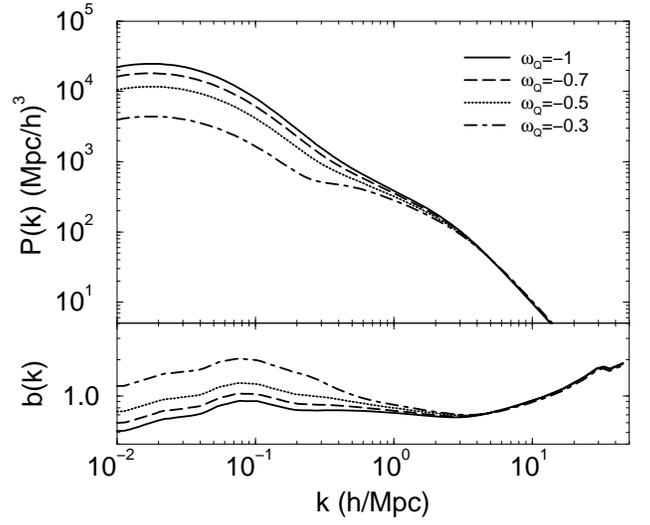}
\epsfxsize=3.6in
\epsfbox{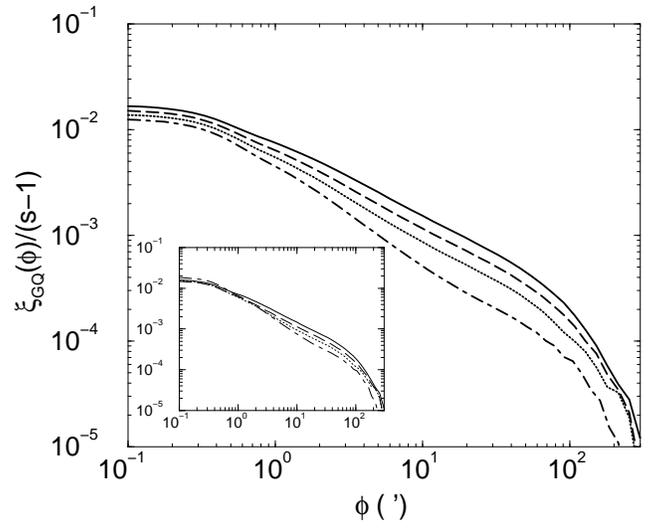}
\caption{Quintessence.
Solid lines are for $\omega_Q=-1$ ($\sigma_8=1.0$); 
dashed ones for $\omega_Q=-0.7$ ($\sigma_8=0.9$); 
dotted ones for $\omega_Q=-0.5$ ($\sigma_8=0.76$); and
dot-dashed ones for $\omega_Q=-0.3$ ($\sigma_8=0.54$).
Other parameters are 
$ (\Omega_m, \Omega_Q,  \Omega_b) 
= (     0.3,      0.7, 0.019/h^2)$, 
$h=0.7$, $n=1$, non-linear approximation by Ma et al. (1999).}
\label{Quintessence}
\end{figure}

Our results are shown in Fig. \ref{Quintessence}. 
A greater $\omega_Q$ (closer to 0) implies a lower amount of
large-scale 
structures over small ones in the relevant range (at much
larger scales the clustering of the quintessence field becomes important),
which results in a steeper slope for $\xi_{GQ}(\phi)$. 
The same conclusion is valid if we normalize the $QCDM$ spectra to the
cluster abundance, which also depends on $\omega_Q$ according to 
Wang \& Steinhardt (1998). 
Note that, because the larger scale structures are the most affected
by the coefficient of the quintessence equation of state, the effect on
the correlation is more dramatic at large angles. 

\subsection{Alternatives for structure formation}

We also investigated alternatives to adiabatic fluctuations as
mechanism for structure formation: isocurvature fluctuations and 
the cosmic strings scenario. 

A general treatment of initial perturbations allows the existence of 
isocurvature modes (Bucher, Moodley, \& Turok 2000). 
We consider an isocurvature CDM model for the mass power spectrum using
the transfer function by Bardeen et al. (1986) and the spectral index 
obtained by Peebles (1999). In this case we use PD96 to obtain the
non-linear spectrum and normalize it to the cluster abundance.

Except for inflationary models, topological defect models are the only known
way of seeding structure formation (see e.g. Vilenkin and Shellard
1994, Hindmarsh and Kibble 1995, and Brandenberger 1994 for recent reviews). 
We explore this scenario for structure formation through the mass
power spectrum generated by a network of cosmic strings (Avelino,
Caldwell \& Martins 1997). 
Note that the formalism used is insensitive to deviations from
Gaussianity. 
Such deviations are a common feature of cosmic strings scenarios. 
Again, the non-linear spectrum is obtained by PD96 prescription and is
CMB-normalized, which implies $\sigma_8 = 0.46$. A normalization to the
cluster abundance requires $\sigma_8 = 0.81$ [see Avelino, Wu \&
Shellard (2000); a somewhat smaller normalization was obtained by van
de Bruck (1999)].  

\begin{figure}
\centering 
\epsfxsize=3.6in
\epsfbox{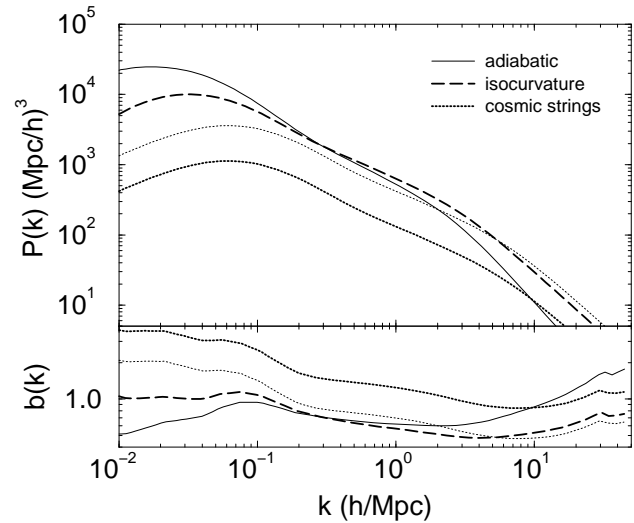}
\epsfxsize=3.6in
\epsfbox{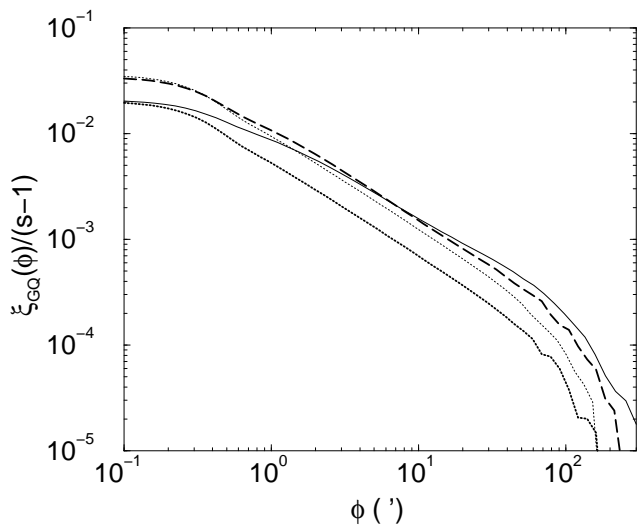}
\caption{Alternative models for structure formation. 
Solid lines are for adiabatic fluctuations ($n=1$, $\sigma_8=1.0$);
dashed ones for isocurvature fluctuations ($n=-1.8$, $\sigma_8=0.99$); and
dotted ones for cosmic strings ($\sigma_8=0.46$). The thin dotted line
corresponds to the normalization of the cosmic strings spectrum to the 
cluster abundance  ($\sigma_8=0.81$).
Other parameters are 
$\Omega_m=0.3$, $\Omega_{\Lambda}=0.7$, $\Omega_b=0.019/h^2$ (cosmic
strings results do not incorporate baryons), $h=0.7$, 
non-linear approximation by PD96.}
\label{alternative}
\end{figure}

Both scenarios are in serious trouble if considered as the only
mechanism responsible for structure formation. However, hybrid models are still
allowed and maybe even favored by the first results from the Boomerang
data (Bernardis et al. 2000) which yields an indication of a low
amplitude for the secondary acoustic peak. 
See Bouchet et al. (2000), Battye \& Weller (2000), and
Avelino, Caldwell \& Martins (1999) for considerations on hybrid models
with topological defects, and Enqvist, Kurki-Suonio \& Valiviita
(2000) for limits on isocurvature fluctuations from CMB anisotropies. 
Our results for these two alternative models for structure formation
are compared with the adiabatic case in Fig. \ref{alternative}. 
Both predict a steeper slope for $\xi_{GQ}(\phi)$ than adiabatic
fluctuations. This is because their power spectra have more
power on smaller scales relative to large ones than the adiabatic
spectrum. 

%%%%%%%%%%%%%%%%%%%%%%%%%%%%%%%%%%%%%%%%%%%%%%%%%%%%%%%%%%%%%%%%%%%%%%%%%%%%%
%%%%%%%%%%%%%%%%%%%%%%%%%%%%%    Discussion     %%%%%%%%%%%%%%%%%%%%%%%%%%%%%
%%%%%%%%%%%%%%%%%%%%%%%%%%%%%%%%%%%%%%%%%%%%%%%%%%%%%%%%%%%%%%%%%%%%%%%%%%%%%

\section{DISCUSSION}
\label{Discussion}

We studied in detail the role of several cosmological parameters and
models of structure formation on the prediction of the angular
correlation between galaxies and QSOs induced by weak gravitational
lensing. 

In a log-log plot $\xi_{GQ}(\phi)$ can be roughly characterized by an 
amplitude and a slope. The amplitude is closely related to the
normalization of the mass power spectrum, and the slope to its shape. 
More power on small scales (large $k$)
implies a steeper slope. This is a reflection of the fact that the amount of 
small scale structure is responsible for the galaxy-QSO correlation on small
angles and larger structures are responsible for the correlation on
larger angles. This simplicity of behavior results in a relatively
large degeneracy of $\xi_{GQ}(\phi)$ as a function of the
cosmological parameters and models analyzed. In brief, a higher
value of $\Omega_m$, $n$, or $h$ has nearly the same effect on 
$\xi_{GQ}(\phi)$ as a lower value of $\Omega_{\nu}$, or $\Omega_b$. 
This effect is a higher correlation on small angles. 
Roughly, this represents a steeper slope for
$\xi_{GQ}(\phi)$, 
which can be also achieved by shifting $\omega_Q$ closer to zero, or by
introducing an isocurvature or cosmic string component.

The main lesson given by the use of different theoretical mass power
spectra is that the galaxy-QSO correlation at different angles
contains information about the amount of structure at different
scales. 

\begin{figure}
\centering 
\epsfxsize=3.6in
\epsfbox{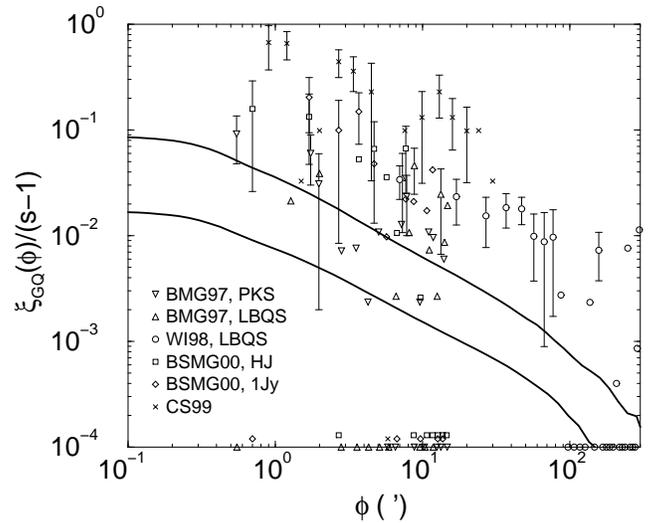}
\caption{Predictions and some observational points.
Solid lines are the theoretical predictions for a flat universe with
cosmological constant and adiabatic fluctuations ($\Omega_m=0.3$,
$\Omega_{\Lambda}=0.7$, $\Omega_b=0.019/h^2$, $n=1$, $h=0.7$,
$\sigma_8=1.0$). The upper curve uses the power spectrum from
Abell-ACO (galaxy clusters), and the lower curve the APM survey
(galaxies).  
Observational points are also plotted as an illustration. Some error
bars would explode in the lower part if plotted, so we do not show
them. 
The points close to the lower axis represent values of
$\xi_{GQ}(\phi)/(s-1)$ equal or less than zero.
Note the caveats listed in the main text concerning the comparison
among different data sets and theoretical predictions.}
\label{observations}
\end{figure}

Our calculations fully incorporate galaxy biasing. This requires the
use of an empirical galaxy power spectrum. We choose to use the
spectrum obtained from the APM survey, but this choice is not
unique. Spectra from other surveys are available and could be used as
well. These spectra do not all agree (see Einasto et al. 1999),
allowing some freedom for the galaxy-QSO
correlation prediction, because our $\xi_{GQ}(\phi)$ calculation can be
seen as an integral over an effective mean spectrum 
$\sqrt{P_{gal}(k)P_{\delta}(k)}$.
In a similar way, the use of a power spectrum of cluster of galaxies
allows for a prediction of the correlation between galaxy clusters and
QSOs.
To illustrate this point we use the spectrum obtained from the
Abell-ACO clusters of galaxies (Miller \& Batuski 2001) as $P_{gal}(k)$.
We show the effect of this alternative choice at
Fig. \ref{observations}, which gives a higher amplitude for
$\xi_{GQ}(\phi)$, roughly given by a factor 
$(\sigma_8)_{Abell-ACO}/(\sigma_8)_{APM}\approx 4$, where
$(\sigma_8)_{Abell-ACO}=3.2$ is the rms fluctuation for the Abell-ACO
spectrum, and $(\sigma_8)_{APM}=0.82$. The correlation curve found
using the Abell-ACO spectrum is also slightly steeper than that using
APM, because the first spectrum is flatter than the second at large
$k$ [$P_{Abell-ACO}(k)\propto k^{-1.2}$ and $P_{APM}(k)\propto k^{-1.4}$].

We also plot in Fig. \ref{observations} some points obtained from
observations. 
A direct comparison among the different data sets is not possible,
because they were obtained from different QSO and galaxy populations.
For the same reason a strict comparison between these observational
points and the theoretical predictions shown is not possible.
That would require an individual calculation for each of the data
sets, which would take in account the different redshift distributions for
QSOs and galaxies, and also the particular power spectrum of the
foreground population (galaxies or galaxy groups).
This would be necessary because when selection criteria are applied to
a foreground object catalog the effective power spectrum of the
resulting subset is possibly different from the spectrum of the whole
population.  
A rigorous analysis of the data would also require considerations about
observational particularities such as the possible effects of dust and
systematics. That is not our aim here. Nevertheless, a qualitative
agreement in behavior between data and predictions can be glimpsed.

Another way to analyze the results from the several groups (avoiding
the messiness of noise data) is to capture the essence of each data
set by adjusting a power law to each one. 
We find the best fit of $\xi_{GQ}(\phi)/(s-1) = A (\phi/1^{\prime})^B$
to the data by the least squares method (see Table \ref{fit-table}). 
To compare with the predictions shown in Fig. \ref{observations} we
can take the amplitude $A$ as being the value of
$\xi_{GQ}(\phi)/(s-1)$ at $\phi=1^{\prime}$ (for the Abell-ACO result 
$A\approx 0.04$ and for APM $A\approx 0.008$), and power index $B$ as the
average bi-log slope of $\xi_{GQ}(\phi)/(s-1)$, which gives $B\approx -0.8$.  
Note however that for  
$\phi {\raise0.3ex\hbox{$\;<$\kern-0.75em\raise-1.1ex\hbox{$\sim\;$}}}
1^{\prime}$ 
the weak lensing assumption is not satisfied (strong lensing becomes
important), so that the predictions for $\xi_{GQ}(\phi)$ are
underestimated in this range.

Ben\'{\i}tez \& Mart\'{\i}nez-Gonz\'alez (1997) (BMG97) found the
galaxy-QSO correlation for angles up to 15 arcmin using radio loud
QSOs from the PKS catalog and also using optically selected
QSOs from the LBQS catalog. 
For this last group there is a large dispersion of
data and points with negative correlation on small angles were found;
the overall result is a null correlation
(dust and selection bias were mentioned as possible explanations). 
The large error bars obtained for the two
quasar samples make these observations compatible with predictions. 
Ben\'{\i}tez et al. (2001) (BSMG00) obtained a
somewhat higher correlation on the same anglular scales using quasars
from the Half-Jansk and 1Jy samples.  

Williams \& Irwin (1998) (WI98) also used a subsample of the LBQS catalog
to determine the galaxy-QSO correlation, but for larger
angles (7 to 300 arcmin). For angles up to nearly 100 arcmin they find
a much higher correlation than the predicted one, and for angles larger
than that the correlation is dispersed around zero, being compatible
with predictions. 
Croom \& Shanks (1999) (CS99) reinterpreted the anti-correlation between
faint QSOs and galaxy groups found by Boyle, Fong \& Shanks (1988) as
being due to gravitational lensing. A flat QSO number count ($s=0.78$) was
used to explain the negative correlation, but a surprisingly high
amplitude is obtained when $\xi_{GQ}(\phi)/(s-1)$ is calculated. 
This high amplitude would be in concordance with a higher
normalization of the power spectrum of the galaxy groups.
Other works also seek to find the observed galaxy-QSO correlation, but
the general conclusion that one can reach is that up to now
$\xi_{GQ}(\phi)$ is an observationally not very well determined quantity. 

The use of much larger catalogs (e.g. the Sloan Digital Sky Survey with
$\approx 10^5$ QSOs, and the 2dF survey) may allow a more precise and
accurate measurement of $\xi_{GQ}(\phi)$ which in turn may constitute
an independent way for the determination of the mass power spectrum,
and a valuable tool for investigating structure formation and
large-scale structure.

\section*{ACKNOWLEDGEMENTS}
ACCG thanks Ian dell'Antonio for his valuable
comments, and also the Rutgers University Physics Department for
the occasional use of its facilities.
C.vdB was supported by the 
Deutsche Forschungsgemeinschaft DFG (at Cambridge) 
and by NATO/DAAD (at Brown). The research at Brown was supported in part
by the US Department of Energy under Contract DE-FG0291ER40688, Task A.

%%%%%%%%%%%%%%%%%%%%%%%%%%%%%%%%%%%%%%%%%%%%%%%%%%%%%%%%%%%%%%%%%%%%%%%%%%%%%
%%%%%%%%%%%%%%%%%%%%%%%%%%%%%%%% references %%%%%%%%%%%%%%%%%%%%%%%%%%%%%%%%
%%%%%%%%%%%%%%%%%%%%%%%%%%%%%%%%%%%%%%%%%%%%%%%%%%%%%%%%%%%%%%%%%%%%%%%%%%%%%

%%%%%%%%%%%%%%%%%%%%%%%%%%%%%%%%%%%%%%%%%%%%%%%%%%%%%%%%%%%%%%%%%%%%%%%%%
%%%%%%%%%%%%%%%%%%%%%%%%%%%%%%%%%%%%%%%%%%%%%%%%%%%%%%%%%%%%%%%%%%%%%%%%%

\newpage

%\protect
\onecolumn
\begin{table}
\caption{Best fit to $\xi_{GQ}(\phi)/(s-1) = A
  (\phi/1^{\prime})^B$, with $B \leq 0$. $\phi$ is the angular range covered
  in arcmin, and $N$ is the number of points. The uncertainties are
  given by the projection of the  $1\sigma$ confidence level contour,
  and $(...)$ represents a large value outside the investigated
  parameter space. Values for $s$ were taken from the respective papers.}
\begin{center}
\begin{tabular}{l r@{.}l c c c r@{.}l r@{.}l}
\hline 
data set       & \multicolumn{2}{c}{$s$}   & $\phi\;(\;^{\prime})$&  N
&  $\chi^2$  &
\multicolumn{2}{c}{$A$}                     & \multicolumn{2}{c}{$B$}\\
\hline
BMG97, PKS     &  3&5  &  0.5--15             & 20  &  17.0      &
0&$05^{+0.03}_{-0.03}$  & -1&$3^{+0.5}_{-(...)}$ \\
BMG97, LBQS    &  2&5  &  0.5--15             & 20  &  19.0      &
0&$00^{+0.03}_{-0.07}$  & 0&$^{+0}_{-(...)}$      \\
WI98, LBQS     &  2&75 &  7--300              & 30  &  40.1      &
0&$3^{+0.3}_{-0.2}$     & -1&$0^{+0.2}_{-0.3}$ \\
WI98, LBQS     &  2&75 &  7--100              & 10  &  3.83      &
0&$11^{+0.14}_{-0.07}$  & -0&$6^{+0.3}_{-0.3}$ \\
BSMG00, HJ     &  1&42 &  0.7--15             & 15  &  11.1      &
0&$1^{+0.1}_{-0.1}$     & -1&$4^{+0.9}_{-(...)}$ \\
BSMG00, 1Jy    &  1&93 &  0.7--15             & 15  &  13.7      &
0&$1^{+0.1}_{-0.1}$     & -1&$^{+1}_{-(...)}$    \\
CS99           &  0&78 &  0.9--30             & 15  &  14.5      &
0&$5^{+0.2}_{-0.2}$     & -0&$6^{+0.2}_{-0.3}$ \\
\hline
\end{tabular}
\end{center}
\label{fit-table}
\end{table}
\twocolumn

\end{document}